\documentclass[%
 reprint,
nofootinbib,
 amsmath,amssymb,
 aps,
]{revtex4-1}

\usepackage{graphicx,setspace}
\usepackage{dcolumn}
\usepackage{bm}
\usepackage[mathlines]{lineno}
\usepackage[colorlinks=true,citecolor=blue,linkcolor=blue,anchorcolor=green, anchorcolor=green,urlcolor=blue]{hyperref}
\usepackage{mathrsfs}
\usepackage{xcolor}
\usepackage{float}
\usepackage{ulem}
\usepackage{lipsum} 

\let\originalleft\left
\let\originalright\right
\renewcommand{\left}{\mathopen{}\mathclose\bgroup\originalleft}
\renewcommand{\right}{\aftergroup\egroup\originalright}
\mathcode`\*="8000
{\catcode`\*=\active\gdef*{\mathclose{}\,\mathopen{}}}

\newcommand{\cu}[1]{\left\{#1\right\}}

\newcommand{\be}{\begin{equation}}
\newcommand{\ee}{\end{equation}}
\newcommand{\bea}{\setlength\arraycolsep{2pt} \begin{eqnarray}}
\newcommand{\eea}{\end{eqnarray}}
\newcommand{\nn}{\nonumber}
\newcommand{\subbe}{\begin{subequations}}
\newcommand{\subee}{\end{subequations}}

\begin{document}

\title{Influence of a plasma on the observational signature of a high-spin Kerr black hole}

\author{Haopeng Yan}
\affiliation{\small
The Niels Bohr Institute, University of Copenhagen, Blegdamsvej 17, 2100 Copenhagen {\O}, Denmark}


\begin{abstract}
  To approach a more reliable observational signature of a high-spin Kerr black hole, one should take into account the effects of its surroundings. To this end we study in this paper the influence of a surrounding plasma. We consider its refractive and dispersive effects on photon trajectories and ignore the gravitational effects of plasma particles as well as the absorption or scattering processes of photons.
  With two specific plasma models, we obtain analytical formulae for the black hole shadow and for the observational quantities of an orbiting ``hot spot" seen by an observer located far away from the black hole.
  We find that the plasma has a frequency-dependent dispersive effect on the size and shape of the black hole shadow and on the image position and redshift of the hot spot. These results may be tested by the Event Horizon Telescope in the future.
\end{abstract}

\maketitle




\section{Introduction}\label{Sec:Introduction}
Nowadays, we are entering an exciting new era of precise astronomical observations of black holes. The observations with gravitational waves have achieved a celebrating breakthrough in recent years \cite{Abbott:2016blz,Abbott:2016nmj,Abbott:2017gyy,Abbott:2017xzg,Abbott:2018oah}. Meanwhile, the Event Horizon Telescope (EHT) collaboration is making efforts on capturing the first image of an astrophysical black hole through electromagnetic wave observations \cite{Falcke:2018uqa}. Therefore, there is increasing interest in studying theoretical templates for those observations among the gravity community \cite{Barack:2018yly,Cunha:2018acu}. The optical signature of a high-spin Kerr black hole at EHT has been studied recently in Refs.~\cite{Gralla:2017ufe,Lupsasca:2017exc,Gates:2018hub}, where the authors found some striking signatures which may serve as `smoking gun' to identify the black hole in the universe. A generalization of these signatures for a Kerr-like black hole in a modified gravity has been discussed in Ref.~\cite{Guo:2018kis}. In these studies, the light rays were assumed as lightlike geodesics of the\linebreak spacetime without being influenced by the medium they passed through. However, an astrophysical black hole is usually surrounded by a complicated environment (such as a corona, a plasma and jets, etc.) and photons near the black hole have to pass through this before reaching to an observer far away from the black hole. In general, the influence of these surroundings on astronomical observations can not be neglected. Then, what about the signatures at EHT if we take this influence into consideration?

Though there are various forms of matter surrounding a black hole, in this work we will only concentrate on the influence of a plasma. Plenty of astronomical phenomena of a black hole in a plasma have been studied ever since the 1960s \cite{muhleman1966radio,muhleman1970radio} while there were also some recent studies on gravitational lensing \cite{BisnovatyiKogan:2010ar,Rogers:2015dla,morozova2013gravitational,Benavides-Gallego:2018ufb,Schee:2017hof,Crisnejo:2018uyn,
Bisnovatyi-Kogan:2017kii}, and shadow of black holes \cite{Perlick:2015vta,Atamurotov:2015nra,Abdujabbarov:2015pqp,Bisnovatyi-Kogan:2017kii,Perlick:2017fio} and wormholes \cite{Abdujabbarov:2016efm}. Here, we are aiming to find the influence of a plasma on: a) the shape and size of shadow for a high-spin black hole, and b) the image of an orbiting emitter (``hot spot") near the black hole. To achieve this target analytically, we will consider several idealized plasma models which have power-law-like distributions and satisfy a separation condition proposed by Perlick and Tsupko \cite{Perlick:2017fio}. The shadow for a Kerr black hole in a plasma has been studied in Refs.~\cite{Atamurotov:2015nra,Perlick:2017fio,Huang:2018rfn}. However, it is worth to revisit this for a high-spin black hole since doing so helps one to understand the image of a hot spot (and thus, the signature at EHT) better. Moreover, in contrast with these works, we will calculate the shadow either using a different method or with different plasma models (or both). The complete signature at EHT should be the combined information of the black hole shadow and the signal from the hot spot. In addition to the influence on the size and shape of a black hole discussed in \cite{Perlick:2017fio,Huang:2018rfn}, we find that there is a special segment of the shadow edge originating from the near-horizon region and is approximately the same for both of the power-law-like models \eqref{eq:PlasmaCase1} and \eqref{eq:PlasmaCase2}. Moreover, the image position and redshift of the hot spot are obviously influenced by the plasma as well. Furthermore, this observational signature is frequency-dependent and there is a greater influence on light rays with lower frequencies.

This paper is organized as follows.
In Sec.~\ref{Sec:PhotonMotion}, we review the photon motion in Kerr spacetime with a plasma and introduce two plasma models with radial power-law-like distributions to be considered later. In Sec.~\ref{Sec:Shadow}, we revisit the shadow of a Kerr black hole in the presence of a plasma, in particular, we study the extremal limit of the shadow and the near horizon extremal Kerr line (NHEKline). In Sec.~\ref{Sec:OrbitingEmitter}, we write down the lens equations for an orbiting emitter and find solutions for a near-extremal black hole to the leading order in the deviation from extremality. In Sec.~\ref{Sec:ObservationalAppearance}, we present the results for the observational appearance of this orbiting emitter and illustrate these with figures. In Sec.~\ref{Sec:Conclusion}, we give a summary and short conclusion. We relegate some technical steps and discussions to appendixes.

\section{Photon motion in Kerr spacetime with a plasma}
\label{Sec:PhotonMotion}
\subsection{Photon motion}
\label{Subsec:PhotonMotion}
We work in the Kerr spacetime which is thought to describe astrophysical black holes in our universe.
The Kerr metric in Boyer-Lindquist coordinates, $x=(t,r,\theta,\phi)$, is given by
\be
\label{eq:metric}
ds^2=-\frac{\Delta \Sigma}{\Xi}dt^2+\frac{\Sigma}{\Delta}dr^2+\Sigma d\theta^2+\frac{\Xi \sin^2 \theta}{\Sigma}(d\phi-W dt)^2,
\ee
where
\begin{subequations}
\bea
\Sigma&=&r^2+a^2\cos^2\theta,\qquad
\Delta=r^2-2Mr+a^2,\\
W&=&\frac{2aMr}{\Xi},\qquad
\Xi=(r^2+a^2)^2-\Delta a^2\sin^2\theta.
\eea
\end{subequations}

We consider that there exists a non-magnetized pressureless plasma with electron frequency \cite{BisnovatyiKogan:2010ar}
\be
\label{eq:PlasmaFrequency}
\omega_p(x)^2=\frac{4\pi e^2}{m_e}N_e(x),
\ee
where $e$ and $m_e$ are the electron charge and mass, respectively, and $N_e$ is the electron number density.
In the geometric optics limit, the Hamiltonian for a photon propagation through this plasma can be written as \cite{BisnovatyiKogan:2010ar}
\be
H(x,p)=\frac{1}{2}\Big(g^{\mu\nu}(x)p_\mu p_\nu+\omega_p(x)^2\Big),
\ee
where $p_{\mu}$ are the components of the four-momentum of the photon and $g^{\mu\nu}$ are the contravariant components of the metric. $p=(p_t,p_r,p_{\theta},p_{\phi})$ are the canonical momentum coordinates.

Note that the plasma has a refractive effect on the photon trajectories and the index of refraction $n(x,\omega)$ is given by
\be
\label{eq:RefIndex}
n(x,\omega)^2=1-\frac{\omega_p(x)^2}{\omega(x)^2},
\ee
where $\omega(x)$ is the photon frequency with respect to the plasma medium. For a photon to be able to propagate through this medium, one should require \be
\label{eq:PropagateCondition}
\omega(x)\geq\omega_p(x).
\ee
For details with regard to the plasma theory, readers may refer to Refs. \cite{BisnovatyiKogan:2010ar,Rogers:2015dla}.

In order to find the equation of motion for photons in the Kerr spacetime with a plasma, we should take care of the plasma frequency.
In the vacuum case $\omega_p(x)=0$, there are four constants of the photon motion: the hamiltonian $H=0$, the total energy $E=-p_t$, the angular momentum $L=p_{\phi}$ and the Carter constant $Q=p^2_{\theta}-\cos^2\theta(a^2p^2_t-p^2_{\phi}\csc^2\theta)$. Provided these constants, the photon trajectories are uniquely determined and one may obtain them by solving the Hamilton-Jacobi (H-J) equation.
However, this is no longer the case in general if there is a non-zero plasma.
For photons propagating through a plasma, the Hamiltonian $H=0$ still holds.
If we assume that the plasma frequency depends\linebreak only on $r$ and $\theta$, then $E=-p_t$ and $L=p_{\phi}$ are still constants of photon motion since $\partial_t H=0$ and $\partial_{\phi} H=0$. For later reference, we introduce $\omega_0$ to denote the photon frequency measured at infinity, then we have $E=\hbar\omega_0$ (hereafter we set $\hbar=1$ for convenience). Next, to make the H-J equation separable, the plasma frequency\linebreak $\omega_p(r,\theta)$ should take the following form \cite{Perlick:2017fio},
\be
\label{eq:SeparationCondition}
\omega_p(r,\theta)^2=\frac{f_r(r)+f_{\theta}(\theta)}{r^2+a^2\cos^2\theta},
\ee
with some functions $f_r(r)$ and $f_{\theta}(\theta)$. Therefore, we can get a generalized separation constant,
\bea
K:&=&p_{\theta}^2+(a\omega_0\sin\theta-L\csc\theta)^2+f_{\theta}(\theta)\nn\\
&=&-\Delta p_r^2+\frac{1}{\Delta}\big[(r^2+a^2)\omega_0-aL\big]^2-f_r(r).
\eea
Follow the convention of Refs.~\cite{porfyriadis2017photon,Gralla:2017ufe}, we define the generalized Carter constant as $Q=K-(L-a E)^2$, the explicit expression is
\be
\label{eq:CarterConstant}
Q=p_{\theta}^2-\cos^2\theta(a^2 \omega_0^2-L^2\csc^2\theta)+f_{\theta}(\theta).
\ee
Note that, if $f_{\theta}(\theta)$ is a constant function, we have $Q-f_{\theta}(\theta)\geq0$ for any photon passing through the equatorial plane since $Q-f_{\theta}(\theta)=p_{\theta}^2\geq0$ when $\theta=\pi/2$.
It is convenient to introduce the following rescaled quantities and functions,
\be
\label{eq:rescale}
\hat{\lambda}=\frac{L}{\omega_0},\qquad
\hat{q}=\frac{\sqrt{Q}}{\omega_0},\qquad
\hat{f}_r=\frac{f_r}{\omega_0^2},\qquad
\hat{f}_{\theta}=\frac{f_{\theta}}{\omega_0^2}.
\ee
In the vacuum case, the trajectory of a photon is independent of its frequency and may be described only by the rescaled quantities $\hat{\lambda}$ and $\hat{q}$ \cite{bardeen1973timelike}. In the presence of a plasma, however, the photon trajectory does depend on the photon frequency and should be described also with an additional variable $\omega_0$ (since the functions $f_r$ and $f_{\theta}$ are not variables of photons). This can be seen from the trajectory equations \eqref{eq:EOM}.

Provided that the surrounding plasma satisfy the separation condition \eqref{eq:SeparationCondition}, one can obtain the equation of motion for photons by using the Hamilton-Jacobi (H-J) method, as follows,
\begin{subequations}
\label{eq:EOM}
\bea
\label{eq:RTheta}
&&-\kern-1.05em\int^{r}\frac{dr}{\pm\sqrt{\mathcal{R}(r)}}=
-\kern-1.05em\int^{\theta}\frac{d\theta}{\pm\sqrt{\Theta(\theta)}},\\
\label{eq:Phi}
\Delta\phi&=&-\kern-1.05em\int^{r}\frac{a(2Mr-a\hat{\lambda})}{\pm\Delta
\sqrt{\mathcal{R}(r)}}dr+
-\kern-1.05em\int^{\theta}
\frac{\hat{\lambda}\csc^2\theta}{\pm\sqrt{\Theta(\theta)}}d\theta,\\
\label{eq:T}
\Delta t&=&-\kern-1.05em\int^{r}\frac{1}{\pm\Delta
\sqrt{\mathcal{R}(r)}}\Big[r^4+a^2\big(r^2+2Mr\big)-
2aMr\hat{\lambda}\Big] dr\nn\\&&+
-\kern-1.05em\int^{\theta}
\frac{a^2\cos^2\theta}{\pm\sqrt{\Theta(\theta)}}d\theta,
\eea
\end{subequations}
where
\bea
\label{eq:RadialPotential}
\mathcal{R}(r)&=&\mathcal{R}_{\text{vac}}(r)-\Delta\hat{f}_r(r),\\
\label{eq:AngularPotential}
\Theta(\theta)&=&\Theta_{\text{vac}}(\theta)-\hat{f}_{\theta}(\theta),
\eea
with
\bea
\mathcal{R}_{\text{vac}}(r)&=&\big(r^2+a^2-a\hat{\lambda}\big)^2
-\Delta\Big[\hat{q}^2+\big(
a-\hat{\lambda}\big)^2\Big],\\
\Theta_{\text{vac}}(\theta)&=&\hat{q}^2+a^2\cos^2\theta
-\hat{\lambda}^2\cot^2\theta.
\eea
The functions $\mathcal{R}(r)$ and $\Theta(\theta)$ are the radial and angular potentionals in a plasma, respectively, while the functions\linebreak $\mathcal{R}_{\text{vac}}(r)$ and $\Theta_{\text{vac}}(\theta)$ are the corresponding potentials in the vacuum case, respectively. Note that the trajectory equations have the same formulas as those in the vacuum case but the corrections are implied in these potentials.
The integrals in these equations are to be evaluated as path integrals along each trajectory, thus, we use slash notations to distinguish these with ordinary integrals. The plus/minus sign in these equations are chosen to be the same as those of the corresponding directions of photon propagation (sign of $dr$ or $d\theta$). The direction is changed every time when the light ray meets a turning point where either $\mathcal{R}(r)$ or $\Theta(\theta)$ vanishes.

To summarize, the plasma contributes an additional term to the Hamiltonian of a photon, which makes the H-J equation non-separable in general. By assuming that the plasma frequency satisfies the separation condition \eqref{eq:SeparationCondition}, the equation of motion for photons can be obtained and the influence of a plasma appears only from the radial potential $\mathcal{R}(r)$ and angular potential $\Theta(\theta)$.
\subsection{Plasma models}\label{Subsec:PlasmaModels}
As the plasma surrounds with a stationary and axisymmetric black hole,
we will consider several specific plasma distributions which depend only on $r$ and $\theta$. The simplest and well-studied model is the radial power-law density \cite{Rogers:2015dla} which depends only on $r$ and satisfies
\be
\label{eq:powerlaw}
\omega_p(r)^2=\frac{4\pi e^2 N_e(r)}{m_e},\qquad
N_e(r)=\frac{N_0}{r^h},
\ee
where $N_0$ is a constant and $h\geq0$.
Unfortunately, this does not satisfy the separation condition \eqref{eq:SeparationCondition}, thus the above mentioned procedure for obtaining photon motion can not be applied. Nevertheless, we may assume the plasma density has an additional $\theta$ dependence such that the separation condition is satisfied, and we make a choice for $f_r(r)$ and $f_{\theta}(\theta)$ in Eq.~\eqref{eq:SeparationCondition} as \cite{Perlick:2017fio}
\be
\label{eq:Powerlawlike}
f_r(r)=Cr^k, \qquad
f_{\theta}(\theta)\geq0,
\ee
where $C>0$ and $0\leq k\leq2$. Since the plasma density is supposed to be negligible at infinity, we are going to consider two models with $k=0$ and $k=1$.

Model 1: A plasma density with $f_r(r)=0$, $f_{\theta}(\theta)=\omega_c^2 M^2$ (or equivalently $f_r(r)=\omega_c^2 M^2$, $f_{\theta}(\theta)=0$ ) such that $\omega^2_p\sim\frac{1}{r^2}$ at large $r$,
\be
\label{eq:PlasmaCase1}
\omega_p(r,\theta)^2=\frac{\omega_c^2 M^2}{r^2+a^2\sin^2\theta}.
\ee

Model 2: A plasma density with $f_r(r)=\omega_c^2 Mr$, $f_{\theta}(\theta)=0$ such that $\omega^2_p\sim\frac{1}{r}$ at large $r$,
\be
\label{eq:PlasmaCase2}
\omega_p(r,\theta)^2=\frac{\omega_c^2 Mr}{r^2+a^2\sin^2\theta}.
\ee

We have introduced a constant $\omega_c$ in these models. For later reference, we will also introduce a rescaled constant, $\hat{\omega}_c=\omega_c/\omega_0$. We name the plasma distributions with the form of Eqs.~\eqref{eq:SeparationCondition} and \eqref{eq:Powerlawlike} the power-law-like models for the reason that the distributions are approximately the same as the power-law models at large $r$.

Since we are interested in the optical appearance of a black hole, we may
expect the existence of a light ray anywhere in the outside of the black hole. As mentioned following \eqref{eq:rescale}, light propagation in a plasma does depend on the photon frequency $\omega_0$. The condition \eqref{eq:PropagateCondition} gives a constraint between the plasma frequency $\omega_p$ and the photon frequency $\omega_0$ \cite{Perlick:2017fio},
\be
\omega_p(r,\theta)^2\leq\omega(r,\theta)^2
=-g_{tt}(r,\theta)^{-1}\omega_0^2,
\ee
where
\be
g_{tt}=1-\frac{2Mr}{r^2+a^2\sin^2\theta}.
\ee
As we have already seen from the Eqs.~\eqref{eq:EOM} that, in the presence of a plasma, the relevant quantity to describe a photon trajectory (and thus the observables) is the ratio of plasma frequency and photon frequency which can be represented by the ratio $\hat{\omega}_c$ for these two models. Note that this ratio may reflect two kinds of different physics. On the one hand, if we consider only the photons with a given frequency, different ratios represent different case studies of plasmas with different densities. On the other hand, for a given plasma distribution, different ratios represent the chromatic effect of the plasma. Later we will study the dependence of the optical appearance on the ratio $\hat{\omega}_c$.

For the models \eqref{eq:PlasmaCase1} and \eqref{eq:PlasmaCase2}, we always have $\omega_p<\omega_c$. Therefore,
the photon trajectory is similar as it would propagate in vacuum spacetime if $\hat{\omega}_c\ll1$ [as from Eq.~\eqref{eq:RefIndex} the refraction index $n\rightarrow1$]. However, if $\hat{\omega}_c\gg1$, it is even impossible for a photon to propagate in the plasma.

Even though the plasma models discussed here are highly idealized, it is possible to extract some approximate effects of a real plasma by using these toy models. Therefore,
we will assume the plasma density satisfies the separation \eqref{eq:SeparationCondition} and mostly focus on the power-law-like models throughout the rest of the paper. For convenience, later we will also use the subscripts $s$ and $o$ to represent the source of photons and the observer, respectively.

\section{Shadow of an extremal Kerr black hole in a plasma}\label{Sec:Shadow}
In Ref.~\cite{Perlick:2017fio}, the Kerr shadow in a plasma has been analytically calculated by using the celestial angles \cite{Grenzebach:2014fha} which is appropriate for any position of the observer. Moreover, it is also firstly shown in \cite{Perlick:2017fio} that the analytical approach based on solving the trajectory equations is possible only for a plasma distribution with the form of \eqref{eq:SeparationCondition}\footnote{Previously, Atamurotov et al.~\cite{Atamurotov:2015nra} have analytically calculated the shadow of a Kerr black hole in a plasma but without taking into consideration the separation condition.} (as was reviewed in Sec.~\ref{Sec:PhotonMotion}). In addition, the shadow also has been numerically performed in Ref.~\cite{Huang:2018rfn}. The numerical approach is, in principle, possible for any distribution of plasma (for example, the power-law form and exponential form have been discussed in \cite{Huang:2018rfn}).

Here, we revisit the shadow for a Kerr black hole in the presence of a plasma by using the impact parameters \cite{bardeen1973timelike} (also refer to as `scree coordinates' in literature, see App.~\ref{App:ScreenCoordinates} for a review) which is appropriate for observers at large distances. In particular, we will study the extremal limit of the shadow and take care of the photons in the near-horizon region, whose images are supposed to appear on a vertical line in the vacuum case, the so-called NHEKline \cite{Gralla:2017ufe}. In Sec. \ref{Sec:OrbitingEmitter}, we will further study the image of an orbiting emitter (``hot spot") in this near-horizon region to seek for more signals related to astronomical observations.

The edge of a shadow corresponds to unstable spherical photon orbits around a black hole, which satisfy
\be
\label{eq:SphericalOrbits}
\mathcal{R}(r)=\mathcal{R}^\prime(r)=0,
\ee
where prime denotes derivative. Solving these equations, we have
\begin{subequations}
\label{eq:PhotonRegion}
\bea
\label{eq:PhotonRegionLambda}
\hat{\lambda}&=&-\frac{M(a^2-r^2)+\Delta r\sqrt{1-\delta}}{a(r-M)},\\
\label{eq:PhotonRegionQ}
\hat{q}&=&\frac{r^{3/2}}{a(r-M)}\bigg[2M\Delta\Big(1+\sqrt{1-\delta}\Big)
-r(r-M)^2\nn\\
&&+(r-2M)\Delta\delta-\frac
{a^2(r-M)^2}{r^3}\hat{f}_r(r)\bigg]^{1/2},
\eea
\end{subequations}
where we have introduced
\be
\delta=\frac{r-M}{2r^2}\hat{f}^{\prime}_r(r).
\ee
Note that in the near-horizon limit $r\rightarrow M$, we have $\delta\rightarrow 0$.
As discussed below \eqref{eq:CarterConstant}, a photon orbit crossing the equatorial plane should satisfy
\be
\label{eq:PhotonRegionCondition}
\hat{q}^2-\hat{f}_{\theta}(\theta)\geq0.
\ee
Plugging \eqref{eq:PhotonRegionQ} into \eqref{eq:PhotonRegionCondition} gives a region of spacetime filled with such spherical photon orbits. We call this region the ``photon region". (Note that here the photon region is slightly different from that in Ref.~\cite{Perlick:2017fio} since we only include those photon orbits which cross the equatorial plane).

We use the `screen coordinates' $(\alpha,\beta)$ [Eq.~\eqref{eq:ScreenCoordinates}] to describe the image on the sky. The edge of a black hole shadow is the curve $(\alpha,\beta)$ with Eq.~\eqref{eq:PhotonRegion} for $\hat{\lambda}$ and $\hat{q}$ being plugged in, as
\subbe
\bea
\alpha(r)&=&-\frac{\hat{\lambda}(r)}{\sin\theta_o},\\
\beta(r)&=&\pm\sqrt{\hat{q}(r)^2+a^2\cos^2\theta_o-\hat{\lambda}(r)^2\cot^2\theta_o
-\hat{f}_{\theta}(\theta_o)}\nn\\
&&=\pm\sqrt{\Theta(\theta_o)},
\eea
\subee
where the parameter $r$ ranges over this photon region \eqref{eq:PhotonRegionCondition} and also makes $\beta$ be real at a desired inclination.
\subsection{Extremal limit and NHEKline}\label{Subsec:NHEKline}
Now we consider the extremal limit following the procedure of Ref.~\cite{Gralla:2017ufe}.
Letting $a\rightarrow M$ in Eqs.~\eqref{eq:PhotonRegion}, we have
\subbe
\label{eq:ExtremalPhoton}
\bea
M\hat{\lambda}&=&M^2+Mr(1+\sqrt{1-\delta})-r^2M\sqrt{1-\delta},\\
M\hat{q}&=&\bigg[r^3\Big(2M(1+\sqrt{1-\delta})-r+(r-2M)\delta\Big)\nn\\
&&-M^2\hat{f}_r(r)\bigg]^{1/2}.
\eea
\subee
Then the condition \eqref{eq:PhotonRegionCondition} on the radius $r$ is expressed explicitly as
\be
\label{eq:PhotonRegionExact}
\frac{r^3}{M^2}\big[2M(1+\sqrt{1-\delta})-r+(r-2M)\delta\big]
\geq \hat{f}_r(r)+\hat{f}_{\theta}(\theta)
\ee
The shadow edge is then obtained by plugging \eqref{eq:ExtremalPhoton} into Eqs.~\eqref{eq:ScreenCoordinates}, as follows
\subbe
\label{eq:ShadowEdge}
\bea
\alpha&=&-\bigg[M+r(1+\sqrt{1-\delta})-\frac{r^2}{M}
\sqrt{1-\delta}\bigg]\csc\theta_o,\\
\beta&=&\pm\bigg[\frac{r^3}{M^2}\Big(2M(1+\sqrt{1-\delta})
-r+(r-2M)\delta\Big)\nn\\
&&+(M^2-\alpha^2)\cos^2\theta_o
-\hat{f}_{\theta}(\theta_o)-\hat{f}_r(r)\bigg]^{1/2}.
\eea
\subee
For different choices of a plasma model and an inclination $\theta_o$ of the observer, the curves given by \eqref{eq:ShadowEdge} may either be closed or open. In case of an open curve, there are two endpoints correspond to $r=M$, (i.e., photons originate from the event horizon). For a radial power-law-like plasma with
\be
\label{eq:PowerlawPlasma}
f_r(r)=M^{2-k}\omega_c^2r^k, (0\leq k\leq2),\quad
f_{\theta}(\theta)=0,
\ee
the endpoints are at [plugging $r=M$ into \eqref{eq:ShadowEdge}],
\subbe
\bea
\alpha_{\text{end}}&=&-2M\csc\theta_o,\\
\beta_{\text{end}}&=&\pm M\sqrt{3+\cos^2\theta_o-4\cot^2\theta_o-\hat{\omega}_c^2}.
\eea\subee
These endpoints exist provided that $\beta_{\text{end}}$ is real, which gives a critical inclination for the observer, $\theta_{\text{crit}}<\theta_o<\pi-\theta_{\text{crit}}$, where
\be
\theta_{\text{crit}}=\arctan\sqrt{\frac{8-\hat{\omega}_c^2
-\sqrt{(12-\hat{\omega}_c^2)(4-\hat{\omega}_c^2)}}
{\sqrt{(12-\hat{\omega}_c^2)(4-\hat{\omega}_c^2)}-6+\hat{\omega}_c^2}}.
\ee
Note that there are no endpoint at all for $\hat{\omega}_c>\sqrt{3}$ and in that case the given curve is closed. Note also that a real $\beta_{\text{end}}$ also requires that \be
\hat{\omega}_c\leq\hat{\omega}_{\text{crit}}=\sqrt{3+\cos^2\theta_o-4\cot^2\theta_o}.
\ee
Later we will consider that the observer is located at a nearly edge-on inclination, $\theta_o=84.27^{\circ}$ (corresponding to $\hat{\omega}_{\text{crit}}\approx\sqrt{2.97}$), thus, for this observer the endpoints exist provided that $\hat{\omega}_c\lesssim\sqrt{2.97}$. For later reference, we refer to the plasma with $\hat{\omega}_c\lesssim\sqrt{2.97}$ as ``low density" plasma, otherwise as ``high density" plasma.

Since the edge of a shadow does close for all $a<M$, such an open curve has missed an important piece originating from the near-horizon sources. To recover the missing part, we consider the extremal limit again by introducing
\be
a=M\sqrt{1-\sigma^2}, \qquad
r=M(1+\sigma R),
\ee
where $\sigma$ is a small parameter. Then for photons orbits which cross the near-horizon region, Eqs.~\eqref{eq:PhotonRegion} give
\be
\hat{\lambda}=2M+\mathcal{O}(\sigma),\quad
\hat{q}=\sqrt{M^2(3-\frac{4}{R})-\hat{f}_r^{(0)}(r)}+\mathcal{O}(\sigma),
\ee
where $\hat{f}_r^{(0)}(r)$ represents the leading order term in $\sigma$. Note that the plasma has no influence on $\hat{\lambda}$ in this limit.
For the radial power-law-like plasma with \eqref{eq:PowerlawPlasma},
we have
\be
\label{eq:PhotonNHEK}
\hat{\lambda}=2M+\mathcal{O}(\sigma),\quad
\hat{q}=M\sqrt{(3-\hat{\omega}_c^2)-\frac{4}{R^2}}+\mathcal{O}(\sigma).
\ee
Then the other piece of shadow edge (originate from near-horizon region) is traced by
\subbe
\label{eq:ShadowNHEK}
\bea
\alpha(R)&=&-2M\csc\theta_o+\mathcal{O}(\sigma),\\
\beta(R)&=&\pm M\sqrt{3+\cos^2\theta_o-4\cot^2\theta_o-\hat{\omega}_c^2-\frac{4}{R^2}}\nn\\
&&+\mathcal{O}(\sigma).
\eea\subee
From the condition \eqref{eq:PhotonRegionCondition} and the requirement $\beta\in\mathbb{R}$, we can get the allowed range of $R$, as
\be
R\in\Big[\frac{2}{\sqrt{3+\cos^2\theta_o-4\cot^2\theta_o-\hat{\omega}_c^2}},
\infty\Big)+\mathcal{O}(\sigma).
\ee
As $\sigma\rightarrow0$ and in the allowed range of $R$, we have
\subbe
\label{eq:NHEKline}
\bea
\alpha&=&-2M\csc\theta_{\theta_o},\\
|\beta|&<&M\sqrt{3+\cos^2\theta_o-4\cot^2\theta_o-\hat{\omega}_c^2}.
\eea
\subee
This gives precisely the missing part of an open curve, which is the generalized NHEKline \cite{Gralla:2017ufe} in the presence of a plasma.
Note that since both of the plasma models \eqref{eq:PlasmaCase1} and \eqref{eq:PlasmaCase2} have the form of \eqref{eq:PowerlawPlasma}, this NHEKline [Eq.~\eqref{eq:NHEKline}] is applicable for both of them and is exactly the same for each particular value of $\hat{\omega}_c$. However, curves given by Eq.~\eqref{eq:ShadowEdge} for these models are different.

To summarize, for an observer at $\theta_o=84.27^{\circ}$, the shadow is given either by Eq.~\eqref{eq:ShadowEdge} for $\hat{\omega}_c>\sqrt{2.97}$ or by the union of Eqs.~\eqref{eq:ShadowEdge} and \eqref{eq:NHEKline} for $0\leq\hat{\omega}_c\lesssim\sqrt{2.97}$.
Note that any near-horizon source in a plasma with the above mentioned models with $\hat{\omega}_c>\sqrt{2.97}$ can not be seen by this observer.

\subsection{Silhouette of black hole}\label{Subsec:ShadowFigure}
We now show the silhouette of a black hole shadow observed at $\theta_o=84.27^{\circ}$ for the two specific plasma models \eqref{eq:PlasmaCase1} and \eqref{eq:PlasmaCase2} in Fig.~\ref{fig:Shadow1}.
For model 1, $\omega_p^2=\omega_c^2 M^2/(r^2+a^2\cos^2\theta)$, we choose $f_r=\omega_c^2M^2$ and $f_{\theta}=0$ (or equivalently, $f_r=0$ and $f_{\theta}=\omega_c^2M^2$); for model 2, $\omega_p^2=\omega_c^2 Mr/(r^2+a^2\cos^2\theta)$, we choose $f_r=\omega_c^2Mr$ and $f_{\theta}=0$.
The silhouettes are obtained by plugging these specific functions $\hat{f}_r$ and $\hat{f}_{\theta}$ into Eq.~\eqref{eq:ShadowEdge} [and Eq.~\eqref{eq:NHEKline}] over the allowed range of $r$ for each given value of $\hat{\omega}_c$. This allowed range can be found numerically from the inequality \eqref{eq:PhotonRegionExact}.

These exhibit the following dependencies of the shadows on the plasma model and on the value of $\hat{\omega}_c$.
Each of the shadow edges for a ``low density" plasma has a vertical part while that for a ``high density" plasma does not.
The shadows shrink in both model 1 and model 2 when $\hat{\omega}_c$ is increased. Moreover, at a given value of $\hat{\omega}_c$, the shadow shrinks more in model 2 than in model 1. This is because the model 2 has a larger plasma density at a given distance $r$ since the density scales like $1/r$ while in model 1 it scales like $1/r^2$.
Furthermore, in both model 1 and 2, photons in the near-horizon region have contributions to the shadows only for $\hat{\omega}_c<\sqrt{2.97}$ (which gives the NHEKlines). At each same value of $\hat{\omega}_c$, the NHEKlines for these two models are the same. When $\hat{\omega}_c$ goes from $0$ to $\sqrt{2.97}$, the NHEKline appears at the same coordinate of $\alpha$ while the maximum absolute value of $\beta$ decreases.

Note that the plasma distributions of example 2 and example 3 in Ref.~\cite{Perlick:2017fio} also have the power-law-like form \eqref{eq:Powerlawlike} with $\omega_p^2\sim r^{-2}$ and $r^{-3/2}$, respectively. The results\linebreak in \cite{Perlick:2017fio} are exhibited with figures for an observer at $r_o=5M$ and $\theta_o=\pi/2$ and for spin $a=0.999M$. While our models 1 and 2 have $\omega_p^2\sim r^{-2}$ and $r^{-1}$, respectively, and the results are obtained for $a\rightarrow M$, $r_o\rightarrow\infty$ and $\theta_o=84.27^{\circ}$.
We find a good agreement between our results and those in \cite{Perlick:2017fio} on the general features discussed above (except details of the NHEKlines since these have not been discussed in \cite{Perlick:2017fio}).
Moreover, the critical ratios for the photon regions are also quantitatively comparable among these results up to factors in these plasma models and the approximations of observers' locations and black hole spins.
Furthermore, these features for the power-law-like plasma models also qualitatively agree with those for the power-law models [with the form of \eqref{eq:powerlaw}] which have been numerically performed in \cite{Huang:2018rfn} for $\omega_p^2\propto r^{-1}$, $r^{-2}$ and $r^{-3}$. Therefore, even through the separation condition \eqref{eq:SeparationCondition} has been proposed based on a mathematical motivation \cite{Perlick:2017fio}, it is nevertheless physically effective.

\begin{widetext}{2}
\begin{figure}[ht!]
\begin{center}
\includegraphics[width=16cm]{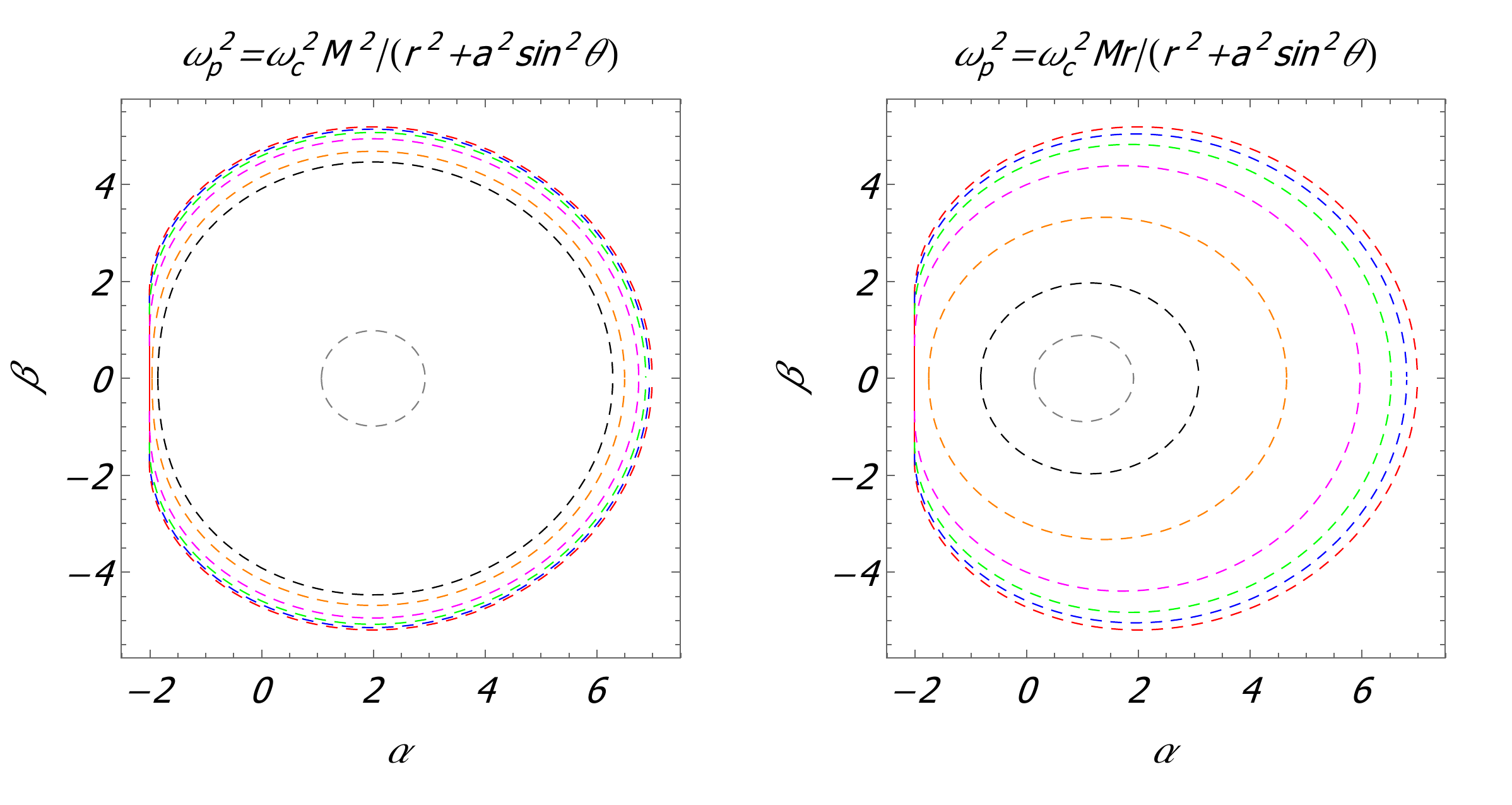}
\end{center}
	\caption{Shadow edge of an extremal Kerr black in the presence of a plasma for model 1 (left) and model 2 (right), respectively, seen from an inclination of $\theta_o=84.27^{\circ}$. For model 1, we have $\omega_p^2=\omega_c^2 M^2/(r^2+a^2\cos^2\theta)$; for model 2, we have $\omega_p^2=\omega_c^2 Mr/(r^2+a^2\cos^2\theta)$. The photon regions for a spherical orbit crossing the equatorial plane vanish at $\hat{\omega}_c^2\approx27$ for model 1 and at $\hat{\omega}_c^2\approx8$ for model 2. For both models, we have $\hat{\omega}_c^2=\omega_c^2/\omega_0^2=0$ (red), 0.5 (blue), 1.2 (green), 2.5 (magenta), 5 (orange) and 7 (black). In addition, the gray curves have $\hat{\omega}_c^2=26$ (left) and 7.8 (right). The dashed lines are given by Eq.~\eqref{eq:ShadowEdge} and the solid vertical lines are given by Eq.~\eqref{eq:NHEKline}. Note that for $\hat{\omega}_c^2=0.5,1.2$ and 2.5, the NHEKlines are overlapped with the red solid lines but have shorter lengths which begin and end at the endpoints of the corresponding dashed lines.}
	\label{fig:Shadow1}
\end{figure}
\end{widetext}
\section{Orbiting emitter in a plasma}\label{Sec:OrbitingEmitter}
Now we consider an isotropic point emitter (``hot spot") orbiting on a circular and equatorial geodesic at radius $r_s$ around a Kerr black hole in the presence of a plasma. This point emitter is supposed to be much heavier than the plasma (for example, it could be a\linebreak star), thus, we may neglect the influence of the plasma on the motion of this emitter.
The angular velocity for such an emitter is given by \cite{bardeen1972rotating} (same as in vacuum Kerr spacetime)
\be
\label{eq:AngularVelocity}
\Omega_s=\pm\frac{M^{1/2}}{r^{3/2}_s\pm aM^{1/2}},
\ee
and the innermost stable circular orbit (ISCO) is given by
\be
\label{eq:Risco}
r_{\text{ISCO}}=M\Big(3+Z_2-\sqrt{(3-Z_1)(3+Z_1+2Z_2)}\Big),
\ee
where
\bea
Z_1&=&1+(1-a^2_{\star})^{1/3}\big[(1+a_{\star})^{1/3}+(1-a_{\star})^{1/3}\big],
\nn\\
Z_2&=&(3a^2_{\star}+Z_1^2)^{1/2},\qquad
a_{\star}=\frac{a}{M}.
\eea

\subsection{Lens equations}\label{Subsec:LensEquations}
The orbiting emitter at $(t_s,r_s,\theta_s,\phi_s)$ is connected to an observer at $(t_o,r_o,\theta_o,\phi_o)$ by photon trajectories described by the equations \eqref{eq:EOM}. The interval of the integrals in these equations are chosen from $r_s$ to $r_o$ and from $\theta_s$ to $\theta_o$, respectively. In addition, we have $\Delta\phi=\phi_o-\phi_s$ and $\Delta t=t_o-t_s$. Using the relation $\phi_s=\Omega_s t_s$, we have
\be
\Delta\phi-\Omega_s\Delta t=\phi_o-\Omega_s t_o,
\ee
then the unknowns $\phi_s$ and $t_s$ can be decoupled from the equations \eqref{eq:Phi} and \eqref{eq:T}.

Following Ref.~\cite{Gralla:2017ufe}, we rearrange these equations \eqref{eq:EOM} as the Kerr lens equations in the presence of a plasma. First, we introduce parameters $b\in\{0,1\}$, $m\in\mathbb{Z}^{\geq0}$ to denote the number of radial and angular turning points, respectively, and $s\in\{-1,1\}$ to denote the final orientation of $p_{\theta}$. Then we set $\phi_o=2\pi l$ with $l\in\mathbb{Z}$ recording the net winding number executed by the photon relative to the emitter between its emission time and reception time.
Finally, the lens equations can be re-expressed as
\begin{subequations}
\label{eq:LensEqns}
\bea
I_r+b\tilde{I}_r&=&G^{m,s}_{\theta},\label{first eq}\\ J_r+b\tilde{J}_r+\frac{\hat{\lambda}G^{m,s}_{\phi}-\Omega_sa^2G^{m,s}_t}
{M}&=&-\Omega_s t_o+2\pi l,\label{second eq}
\eea
\end{subequations}
where $I_r$, $\tilde{I}_r$, $J_r$ and $\tilde{J}_r$ are radial integrals defined in appendix \ref{App:RadialIntegrals} and $G^{m,s}_i$ ($i\in\{t,\theta,\phi\}$) are angular integrals defined in appendix \ref{App:AngularIntegrals}.
These equations have the same formulae as those in vacuum Kerr spacetime, however, the differences are implied in the integrals.
Solving these lens equations for given parameters $m,s,b$ and given values of $r_s$ and $\theta_o$, we can write the conserved quantities $\hat{\lambda}$ and $\hat{q}$ in terms of $t_o$ which label the photon trajectories connecting the source to an observer.
\subsection{Near-extremal solutions}\label{Subsec:NearExtremalSolution}
We assume the emitter is on, or near, the prograde ISCO of a near-extremal black hole.
It is convenient to work with a dimensionless radial coordinate $R$ which is related to the Boyer-Lindquist radius $r$ by
\be
\label{eq:DimensionlessR}
R=\frac{r-M}{M}.
\ee
We introduce a small parameter $\epsilon\ll1$ to describe the condition for the near-extremality of a black hole, as follows,
\be
\label{eq:NearExtremalCondition}
a=M\sqrt{1-\epsilon^3}.
\ee
Under this condition, the ISCO [Eq.~\eqref{eq:Risco}] is located at a coordinate distance $\sim\epsilon$ from the horizon,
\be
R_{\text{ISCO}}=2^{1/3}\epsilon+\mathcal{O}(\epsilon^2),
\ee
thus, the radial coordinate for the emitter can be written as
\be
\label{eq:SourcePosition}
r_s=M(1+\epsilon \bar{R})+\mathcal{O}(\epsilon^2),\qquad
\bar{R}\geq\frac{1}{2},
\ee
which means that the emitter is in the near-horizon region. Even though the motion of this emitter is not affected by the plasma, its image can only possibly be seen if the plasma density has $\hat{\omega}_c\lesssim\sqrt{3}$ as the light rays are refracted by the plasma, and in that case the image appears on the NHEKline (see discussions in Sec.~\ref{Subsec:ObsersvationalQuantities}).

Following Ref.~\cite{porfyriadis2017photon,Gralla:2017ufe}, we introduce new quantities $\lambda$ and $q$ instead of $\hat{\lambda}$ and $\hat{q}$ (see appendix \ref{App:Interpretation} for interpretations).
\be
\label{eq:Newquantities}
\hat{\lambda}=2M(1+\epsilon\lambda),\qquad
\hat{q}=M\sqrt{3-q^2}.
\ee

Plugging the expressions \eqref{eq:NearExtremalCondition}, \eqref{eq:SourcePosition} and \eqref{eq:Newquantities} into the lens equations \eqref{eq:LensEqns} gives the near-extremal lens equations.
Then for given values of $r_s$ and $\theta_o$, we solve these equations to the leading order in $\epsilon$ for the plasmas with distributions \eqref{eq:PlasmaCase1} and \eqref{eq:PlasmaCase2}, respectively, following the procedure of Ref.~\cite{Gralla:2017ufe}.

Note that even though we have considered a near-extremal black hole with the condition \eqref{eq:NearExtremalCondition}, it turns out that the results are identical to the leading order in $\epsilon$ if we considered an extremal black hole with $a=M$ \cite{Gralla:2017ufe}.

\subsubsection{Model 1: $\omega_p^2=\omega_c^2 M^2/(r^2+a^2\sin^2\theta)$}
We choose $f_r(r)=0$ and $f_{\theta}(\theta)=\omega_c^2 M^2$. In this case, the radial potential is the same as that in the vacuum Kerr case \cite{Gralla:2017ufe}, thus, we have the same radial integrals and then we obtain the same formulae of solutions for the lens equations up to corrections in the angular integrals. These integrals are performed in App.~\ref{App:Integrals}.

First, we make a choice of the parameters $b$, $m$ and $s$. Then, from the first equation \eqref{first eq} we can obtain the condition for the existence of a solution,
\begin{subequations}
\label{eq:qRange}
\bea
\bar{R}&<&\frac{4\Upsilon}{q^2}\Bigg(1+\frac{2}{\sqrt{4-q^2}}\Bigg)
\quad \text{if}\quad b=0 ,\\
\bar{R}&>&\frac{4\Upsilon}{q^2}\Bigg(1+\frac{2}{\sqrt{4-q^2}}\Bigg)
\quad \text{if}\quad b=1 ,
\eea
\end{subequations}
and the solution, as follows,
\be
\label{lambda of q}
\lambda=\frac{2\Upsilon}{4-q^2}\Bigg[2- q\sqrt{1+\frac{4-q^2}{2\Upsilon}\bar{R}}\Bigg]  .
\ee
Here, $\Upsilon >0$ is defined by
\be
\label{eq:Upsilon}
\Upsilon\equiv\frac{q^4R_o}{q^2+2R_o+qD_o}e^{-qG^{\bar{m},s}_{\theta}},
\ee
where
\be
D_o=\sqrt{q^2+4R_o+R_o^2},
\ee
and $G^{\bar{m},s}_{\theta}$ is defined in Eq.~\eqref{eq:AngleIntegralMS}, with
\be
\label{eq:MBar}
\bar{m}=m+\frac{1}{qG_{\theta}}\log\epsilon.
\ee

Next, regarding the second equation \eqref{second eq}, we introduce a dimensionless time coordinate $\hat{t}_o$ which is restricted to unit period of the emitter,
\be
\hat{t}_o=\frac{t_o}{T_s}=\frac{\Omega_st_o}{2\pi}=\frac{t_o}{4\pi M}+\mathcal{O}(\epsilon).
\ee
Thus, the second equation can be written in terms of this dimensionless time coordinate, as
\be
\label{eq:SolSecondEq}
\hat{t}_o=
l-\frac{1}{2\pi}\Big(J_r+b\tilde{J}_r+2G^{m,s}_{\phi}
-\frac{1}{2}G^{m,s}_{t}\Big),
\ee
where $J$ integrals and $G$ integrals are given in App.~\ref{App:Integrals}.

Note that we have already obtained a function $\lambda(q)$ [Eq.~\eqref{lambda of q}] in the allowed range of $q$ [Eq.~\eqref{eq:qRange}] from the first equation, the second then gives a function $\hat{t}_o(q)$. Inverting this function in each monotonic domain gives a function $q_i(\hat{t}_o)$ for a given choice of integer $l$. Here we have introduced a discrete integer $i$ to label the monotonic parts of $\hat{t}_o(q)$ in each of the allowed ranges of $q$. Since the observational quantities can be expressed in terms of the conserved quantities $\lambda$ and $q$, each $q_i(\hat{t}_o)$ corresponds to a specific track segment of the emitter's image which can be labeled by $(m,b,s,l,i)$.

\subsubsection{Model 2: $\omega_p^2=\omega_c^2 Mr/(r^2+a^2\sin^2\theta)$}
We choose $f_r(r)=\omega_c^2 Mr$ and $f_{\theta}(\theta)=0$. Since the lens equations for this case are similar to those for model 1 (see App.~\ref{App:Integrals} for details), we can find the solutions in a similar way for a given choice of parameters $b$, $m$ and $l$. For convenience, we introduce a new parameter $\tilde{q}=\sqrt{q^2-\hat{\omega}_c^2}$.

The solution of the first equation \eqref{first eq} and the condition of its existence are given by replacing $q$ with $\tilde{q}$ in the formulas \eqref{lambda of q} and \eqref{eq:qRange}, respectively, where the expression of $\Upsilon$ is corrected as
\be
\Upsilon\equiv\frac{\tilde{q}^4R_o}{\tilde{q}^2+(2-\frac{\hat{\omega}_c^2}{2})
R_o+\tilde{q}D_o}e^{-\tilde{q}
G^{\bar{m},s}_{\theta}},
\ee
with
\be
D_o=\sqrt{\tilde{q}^2+(4-\hat{\omega}_c^2)R_o+R_o^2},
\ee
and $G^{\bar{m},s}_{\theta}$ is given in Eq.~\eqref{eq:AngleIntegralMS} with $\bar{m}$ being defined in \eqref{eq:MBar}.

The second equation \eqref{second eq} can also be rewritten in a form as \eqref{eq:SolSecondEq}, however, the $J$ integrals and $G$ integrals therein are different from those for model 1 (see App.~\ref{App:Integrals}).

Similarly, we can also obtain the image track segment $q_i(\hat{t}_o)$ of the emitter labeled by $(m,b,s,l,i)$.



\section{Observational appearance of the orbiting emitter}
\label{Sec:ObservationalAppearance}
In the vacuum case, the image of an emitter orbiting on the ISCO of a rapidly rotating black hole appearing on the NHEKline has a rich structure \cite{Gralla:2017ufe}.
Next, we will study the influence of plasma on this image.
\subsection{Observational quantities}\label{Subsec:ObsersvationalQuantities}
From the previous section, the photon conserved quantities $q(\hat{t}_o)$ and $\lambda[q(\hat{t}_o)]$ (along trajectories) are obtained for the plasma models \eqref{eq:PlasmaCase1} and \eqref{eq:PlasmaCase2}. These conserved quantities help to describe the observational appearance of the emitter: the image position, redshift and flux. We briefly review this for a general black hole in App.~\ref{App:ObservationalAppearance} and give the results for a near-extremal black hole below.

In the near-extremal limit and to the leading order in $\epsilon$, we have (see Sec.~\ref{Subsec:NearExtremalSolution})
\subbe
\label{eq:NearExtremalExpansion}
\bea
a&=&M,\qquad\qquad\quad
r_s=M(1+\epsilon\bar{R}),\\
\hat{\lambda}&=&2M(1+\epsilon\lambda),\qquad
\hat{q}=M\sqrt{3-q^2}.
\eea
\subee
Then, the apparent position \eqref{eq:ScreenCoordinates} on the celestial sphere is expanded as
\begin{subequations}
\label{ImagePosition}
\bea
\label{ImagePositionX}
\alpha&=&-2M\csc\theta_o,\\
\label{ImagePositionY}
\beta&=&sM\Big(3-q^2+\cos^2\theta_o-4\cot^2\theta_o-
\frac{\hat{f}_{\theta}(\theta_o)}{M^2}\Big)^{1/2},
\eea
\end{subequations}
where $s$ is the final orientation of $p_{\theta}$.
Note that $\lambda$ does not appear to the leading order and $q$ should be in a range such that $\beta$ is real. For an observer at $\theta_o=84.27^{\circ}$, this range is obtained as $0\leq q\lesssim\sqrt{2.97-\hat{f}_{\theta}(\theta_o)/M^2}$. Combining with another range of $q$ discussed in App.~\ref{App:Interpretation}, we find that the image of a hot spot in a plasma with models 1 and 2 appears on the NHEKline (see Sec.~\ref{Subsec:NHEKline}).
The redshift factor \eqref{eq:RedshiftGeneral} is expanded as
\be
\label{eq:ImageRedshift}
g=\frac{1}{\sqrt{3}+\frac{4}{\sqrt{3}}\frac{\lambda}{\bar{R}}}
+\mathcal{O}(\epsilon).
\ee
The flux \eqref{eq:FluxGeneral} is expanded as
\bea
\label{eq:FluxExpansion}
\frac{F_o}{F_N}&=&\frac{\sqrt{3}\epsilon\bar{R}}{2D_s}\frac{qg^3}{\sin\theta_o
\sqrt{\Theta_0(\theta_o)}\sqrt{3-q^2-\frac{\hat{f}_{\theta}(\theta_s)}{M^2}}}
\nn\\
&&\times\Bigg|\det\frac{\partial(B,A)}{\partial(\lambda,q)}\Bigg|^{-1},
\eea
where $A$ and $B$ are defined in \eqref{eq:DefAB}, and
\be
\Theta_0(\theta_o)=3-q^2+\cos^2\theta_o-4\cot^2\theta_o-\frac{\hat{f}_
{\theta}(\theta_o)}{M^2},
\ee
and
\be
D_s=\sqrt{q^2\bar{R}^2-8\lambda\bar{R}+4\lambda^2-\frac{\hat{f}^{(0)}
_r(r_s)}{M^2}\bar{R}^2},
\ee
with $\hat{f}^{(0)}_r(r_s)$ being the leading order term of $\hat{f}_r(r_s)$ in $\epsilon$.

Note that these results are obtained for each given choice of discrete parameters $m$, $s$, $b$ and $l$, which corresponds to a specific image track. The full time-dependent image is completed by finding all such tracks for all choices of these parameters.
The influences of plasma on these observables are introduced from the functions $\hat{f}_r$ and $\hat{f}_{\theta}$, as well as from the quantities $\lambda$ and $q$ which label different photon trajectories.

\subsection{Hot spot image}
\label{subsec:HotSpotImage}
We now describe the observational quantities of the hot spot's image with figures following the procedure of Ref.~\cite{Gralla:2017ufe} and using the open numerical code therein.
The image depends on the choice of the plasma distribution and the parameters $R_o$, $\theta_o$, $\epsilon$ and $\bar{R}$.
We will consider the two plasma models \eqref{eq:PlasmaCase1} and \eqref{eq:PlasmaCase2} with several certain values of $\hat{\omega}_c$, respectively.
In order to compare the results with those for vacuum case \cite{Gralla:2017ufe}, we make the following choice for these parameters:
\begin{subequations}
  \bea
  R_o&=&100,\qquad
  \theta_o=\frac{\pi}{2}-\frac{1}{10}=84.27^{\circ},\\
  \epsilon&=&10^{-2},\qquad
  \bar{R}=\bar{R}_{\text{ISCO}}=2^{1/3}.
  \eea
\end{subequations}

As described in Sec.~\ref{Subsec:NearExtremalSolution}, for each choice of discrete parameters $m$, $b$, $s$, $l$ and an additional label $i$, we can obtain an image track segment $q(\hat{t}_o)$ [and $\lambda(\hat{t}_o)$]. The main observables for the segment are given in Eqs.~\eqref{ImagePosition}, \eqref{eq:ImageRedshift} and \eqref{eq:FluxExpansion}. The completed information of the image is built up by including all such choices of parameters (in practice, we consider only a few values of $m$ and $l$ since the image for others are vanishingly small) \cite{Gralla:2017ufe}.
Below we show the brightest few images for model 1 [Eq.~\eqref{eq:PlasmaCase1}] in Fig.~\ref{fig:Hotspot1} and for model 2 [Eq.~\eqref{eq:PlasmaCase2}] in Fig.~\ref{fig:Hotspot2}, respectively. We
consider four different values of the ratio $\hat{\omega}_c$ for each of the models and also colour-code continuous image tracks in each of these plots.

Comparing Fig.~\ref{fig:Hotspot1} with Fig.~\ref{fig:Hotspot2}, we find that the images of model 1 and model 2 (with a given value of $\hat{\omega}_c$) are very similar. This is because the difference between the lens\linebreak equations \eqref{eq:LensEqns} among these two models is negligibly\linebreak small. Firstly, the lens equations in the near horizon region are the same to the leading order in $\epsilon$ for both\linebreak models. Secondly, even though there are differences appearing in the far region, these have a negligible influence on the image since the plasma densities decrease with $r$ in an inverse power-law behavior. [See the results\linebreak and discussions for plasma distributions with $f_r(r)=M^{2-k}\omega_c^2r^k$, ($k=0,1)$, and $f_{\theta}(\theta)=0$ in App.~\ref{App:Integrals}.] Therefore, here we will only discuss the features exhibited in Fig.~\ref{fig:Hotspot1} for the model 1.

Fig.~\ref{fig:Hotspot1} shows the main observables in a plasma with $\omega_p^2=\omega_c^2 M^2/(r^2+a^2\sin^2\theta)$, where we have taken four different values for $\hat{\omega}_c^2=\omega_c^2/\omega_0^2$ as $0$, $0.5$, $1.2$ and $2.5$. In each case, the green line is for the brightest primary image while others are for the secondary images. Note that the secondary images are, in general, much fainter than the primary image and are important only when different image tracks intersect. Therefore, below we will focus on the feature of the primary image. In each of these cases, the primary image (if any) appears near the center of the NHEKline before moving downward while spiking in brightness. The image appears periodically and the period stays unchanged when $\hat{\omega}_c$ is increased. For $\hat{\omega}_c^2=0$, this corresponds to the vacuum case and the results are agree with Ref.~\cite{Gralla:2017ufe}. For a non-zero plasma, there are remarkable influences on the image position and redshift while smaller influence on the image flux. When $\hat{\omega}_c$ is increased from zero, not only the maximum elevation of the NHEKline ($\beta_{max}$) decreases but also the relative portion of the NHEKline on which appears the image ($\beta/\beta_{max}$) decreases, and so does the redshift factor and time duration of the image. Note that for smaller values of $\hat{\omega}_c$ the primary image is blueshifted while for larger ones it becomes redshifted. The primary image vanishes when $\hat{\omega}_c$ is greater than a critical value and the entire images vanish when $\hat{\omega}_c\gtrsim\sqrt{2.97}$.

\begin{widetext}{2}
\begin{figure}[ht!]
\begin{center}
\includegraphics[width=16cm]{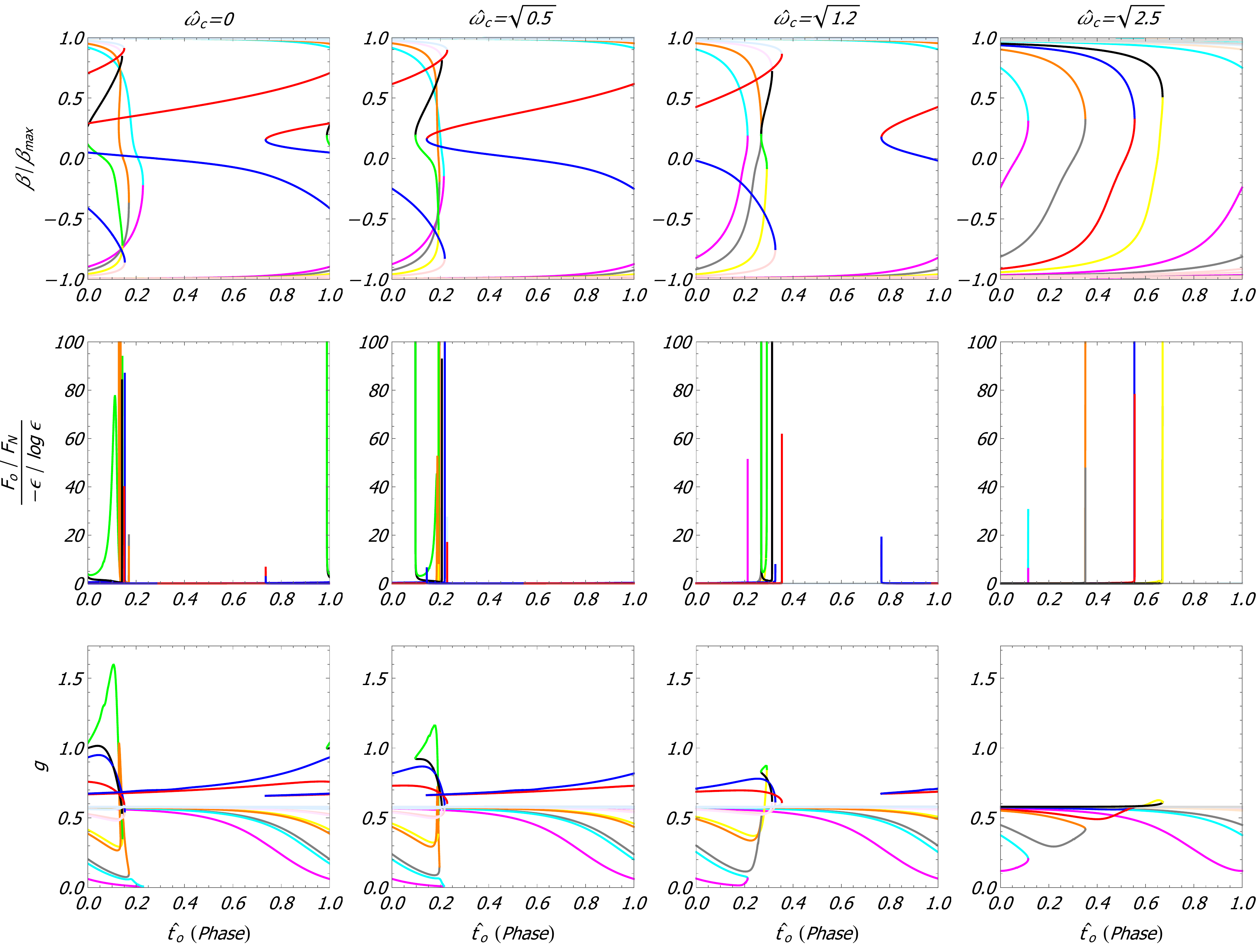}
\end{center}
	\caption{Positions, fluxes and redshifts of the brightest few images of the hot spot for $\omega_p^2=\omega_c^2 M^2/(r^2+a^2\cos^2\theta)$ (model 1). From left to right, we have $\omega_c/\omega_0=0$, $\sqrt{0.5}$, $\sqrt{1.2}$ and $\sqrt{2.5}$, respectively. We have color-coded the images in the same way as that of Ref.~\cite{Gralla:2017ufe} and each monochromatic line may be composed of several continuous track segments labeled by $(m,b,s,l,i)$.}
	\label{fig:Hotspot1}
\end{figure}

\begin{figure}
\begin{center}
\includegraphics[width=16cm]{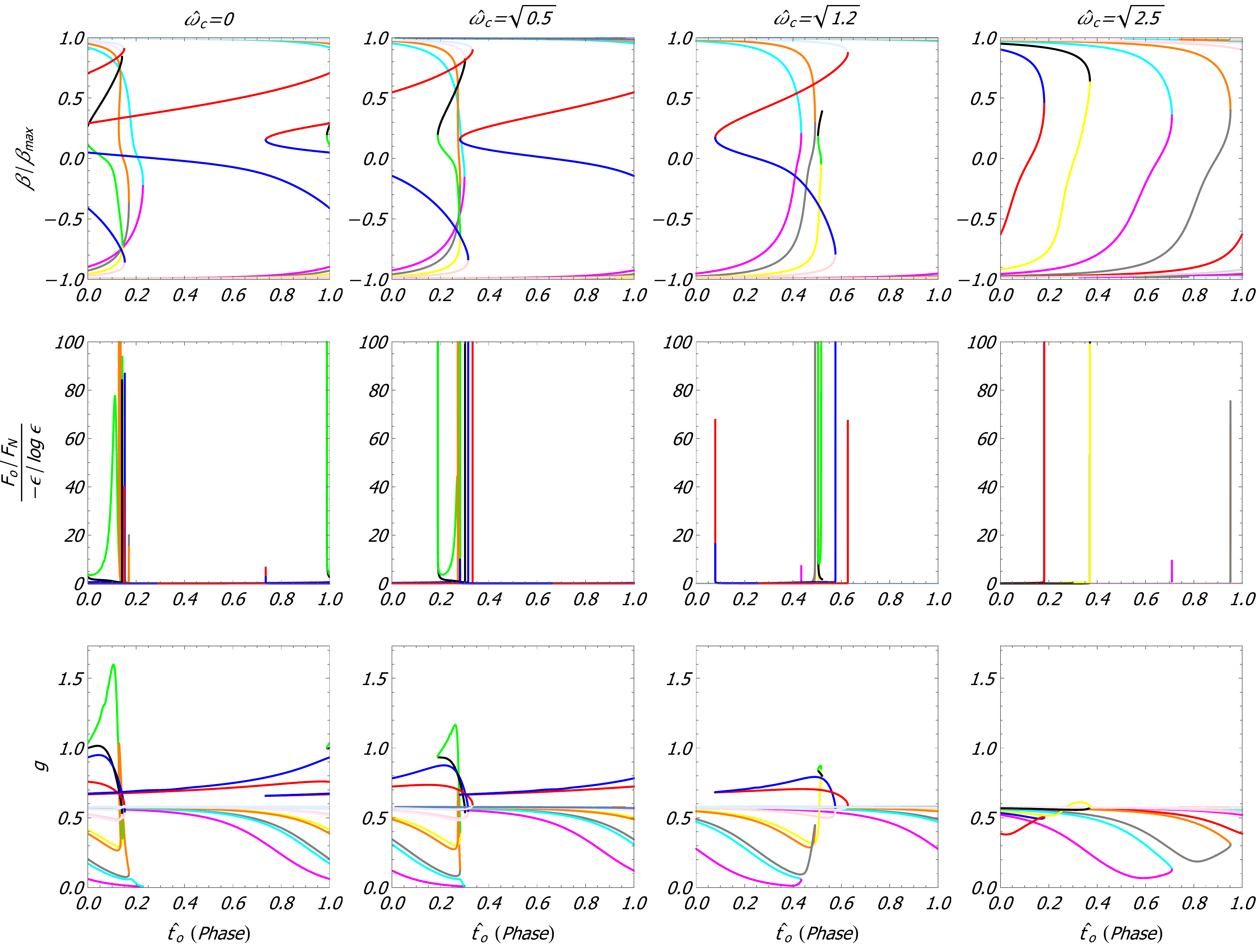}
\end{center}
	\caption{Positions, fluxes and redshifts of the brightest few images of the hot spot for $\omega_p^2=\omega_c^2 Mr/(r^2+a^2\cos^2\theta)$ (model 2). From left to right, we have $\omega_c/\omega_0=0$, $\sqrt{0.5}$, $\sqrt{1.2}$ and $\sqrt{2.5}$, respectively. We have color-coded the images in the same way as that of Ref.~\cite{Gralla:2017ufe} and each monochromatic line may be composed of several continuous track segments labeled by $(m,b,s,l,i)$. 
}
	\label{fig:Hotspot2}
\end{figure}
\end{widetext}

\section{Summary and Conclusion}\label{Sec:Conclusion}
In this paper, we investigated the observational signature of a high-spin black hole in the presence of a surrounding plasma. We consider the plasma as a dispersive medium for photons but neglect its gravitational effects.\linebreak We assume the plasma distributions satisfy a separation condition \eqref{eq:SeparationCondition} proposed by Perlick and Tsupko \cite{Perlick:2017fio} such that the photon trajectory can be solved analytically. Then we study the shadow of the black hole and the signal produced by a nearby hot spot.

To obtain the optical appearance, we first find the\linebreak equation of motion for photons by solving the Hamilton-Jacobi equations under the separation condition. We find that the corrections of these to the vacuum case are imposed only from the radial potential $\mathcal{R}(r)$ and angular potential $\Theta(\theta)$ [see Eq.~\eqref{eq:EOM}]. We also introduce two special power-law-like models [\eqref{eq:PlasmaCase1} and \eqref{eq:PlasmaCase2}] which satisfy the separation condition in Sec.~\ref{Subsec:PlasmaModels} as simple examples which were studied in detail.

Next, we have analytically studied the photon region and shadow of an extremal Kerr black hole surrounded by a plasma. For a power-law-like plasma, the photon region is determined by \eqref{eq:PhotonRegionExact} and the edge of a shadow is described either by the union of Eqs.~\eqref{eq:ShadowEdge} and \eqref{eq:ShadowNHEK} or by Eq.~\eqref{eq:ShadowEdge} only, depending on whether the near-horizon source is in the photon region (``low" density plasma) or not (``high" density plasma). The size of shadow decreases when the density of plasma is increased and the shape of shadow is different for plasma with ``low" or ``high" density. Moreover, in case of ``low" density plasma, the near-horizon sources all terminate at a vertical line: the NHEKline. We show these in Fig.~\ref{fig:Shadow1} and discuss the features in Sec.~\ref{Subsec:ShadowFigure}.

Then we have studied the signal produced by a hot spot orbiting at the ISCO of the black hole (in the presence of a power-law-like plasma). We solve the lens equations in the near-extremal limit in Sec.~\ref{Sec:OrbitingEmitter} and obtain analytical formulae for the observational quantities: the image position \eqref{ImagePosition}, the image redshift \eqref{eq:ImageRedshift} and the image flux \eqref{eq:FluxExpansion}. Note that since the ISCO of a high-spin black hole is in the near-horizon region, this signal can only possible be seen for a ``low" density plasma. The plasma has a remarkable influence on the brightest image: the segment on the celestial sphere for the image to appear is smaller than that in the vacuum case and so does the redshift. We show these in Fig.~\ref{fig:Hotspot1} and \ref{fig:Hotspot2} and discuss the features in Sec.~\ref{subsec:HotSpotImage}.

In a real astrophysical setup (given the parameters for the black hole and the surrounding plasma), the ratio $\hat{\omega}_c$ depends on the frequency of photon, thus, the observational signatures that we discussed above are all chromatic. Note that the ratio $\hat{\omega}_c$ is greater for a photon with lower frequency, thus, there is a larger influence on the trajectory of such photon.
Combining the information of black hole shadow and hot spot signal, we can sketch a picture (template) for what we may see from the Event Horizon Telescope.

\subsection*{Acknowledgements}
The author thanks Minyong Guo and Niels A. Obers \linebreak for helpful discussions and for comments on the manuscript.
The author would like to thank the anonymous referee for the instructive comments.
H. Y. thanks the Theoretical Particle Physics and Cosmology section at the Niels Bohr Institute for support. H. Y. is also financially supported by the China Scholarship Council.

\begin{appendix}
\section{Screen coordinates}
\label{App:ScreenCoordinates}
In this appendix, we introduce a pair of screen coordinates, $\alpha$ and $\beta$, in the rest frame of an observer to describe specific positions on the celestial sphere \cite{bardeen1973timelike}.
We choose the following tetrad for the observer located at $r=r_o$,
\begin{subequations}
\bea
e_{(t_o)}&=&\sqrt{\frac{\Xi}{\Delta\Sigma}}\partial_t
+\frac{2Mar}{\Xi\Sigma\Delta}\partial_{\phi},\quad
e_{(r_o)}=\sqrt{\frac{\Delta}{\Sigma}}\partial_r,\\
e_{(\theta_o)}&=&\frac{1}{\sqrt{\Sigma}}\partial_{\theta},\qquad
e_{(\phi_o)}=\sqrt{\frac{\Sigma}{\Xi}}\frac{1}{\sin\theta}\partial_{\phi}.
\eea
\end{subequations}
The frame components of four vectors are defined in the usual way,
\be
\label{eq:FrameComp}
V^{(a)}=\eta^{(a)(b)}e^{\mu}_{(b)}V_{\mu},
\ee
where $\eta^{(a)(b)}=\text{diag}(-1,1,1,1)$.
For photons with conserved quantities $\hat{\lambda}$ and $\hat{q}$ that reach to the observer, tracing backward along their trajectories to the source gives the image of the source. Thus, the coordinates can be defined with the help of photon motions (see Sec.~\ref{Subsec:PhotonMotion}), as
\subbe
\bea
\label{ScreenCoordinates}
\alpha&=&-r_o\frac{p^{(\phi_o)}}{p^{(t_o)}}=-r_o\frac{\sqrt{\Delta_o}\Sigma_o
\hat{\lambda}\csc\theta_o}{\Xi_o-2aMr_o\hat{\lambda}},\\
\beta&=&\pm r_o\frac{p^{(\theta_o)}}{p^{(t_o)}}=\pm r_o\frac{\sqrt{\Delta_o\Xi_o
\Theta(\theta_o)}}
{\Xi_o-2aMr_o\hat{\lambda}}.\qquad
\eea
\subee
These describe the apparent displacements of the image relative to the center of the black hole: $\alpha$ and $\beta$ are, respectively, in the direction perpendicular/parallel to the axis of symmetry of the black hole.

As $r_o\rightarrow \infty$, the coordinates read
\subbe
\label{eq:ScreenCoordinates}
\bea
\alpha&=&-\frac{\hat{\lambda}}{\sin\theta_o},\\
\beta&=&\pm\sqrt{\hat{q}^2+a^2\cos^2\theta_o-\hat{\lambda}^2\cot^2\theta_o
-\hat{f}_{\theta}(\theta_o)}\nn\\
&&=\pm\sqrt{\Theta(\theta_o)}.
\eea
\subee
\section{Observational appearance of the orbiting emitter}
\label{App:ObservationalAppearance}
\subsection{Interpretation of photon conserved quantities}
\label{App:Interpretation}
We consider the photons originating from the orbiting emitter described in Sec.~\ref{Sec:OrbitingEmitter}. The conserved quantities $\hat{\lambda}$ and $\hat{q}$ are related to the emission angle in the rest frame of the source.
We choose the following tetrad for the emitting source,
\begin{subequations}
\label{eq:EmitterFrame}
\bea
e_{(t_s)}&=&\gamma\sqrt{\frac{\Xi}{\Delta\Sigma}}(\partial_t
+\Omega_s\partial_{\phi}),\\
e_{(r_s)}&=&\sqrt{\frac{\Delta}{\Sigma}}\partial_r,\quad
e_{(\theta_s)}=\frac{1}{\sqrt{\Sigma}}\partial_{\theta}\label{fram 1},\\
e_{(\phi_s)}&=&\gamma v_s \sqrt{\frac{\Xi}{\Delta\Sigma}}(\partial_t+W \partial_{\phi})+\gamma\sqrt{\frac{\Sigma}{\Xi}}\partial_{\phi},
\eea
\end{subequations}
where
\be
v_s=\frac{\Xi_s}{\Sigma_s\sqrt{\Delta_s}}(\Omega_s-W_s),\qquad
\gamma=\frac{1}{\sqrt{1-v^2_s}}.
\ee
The cosines of the emission angles $(\Phi_s,\Theta_s)$ are given by
\bea
\cos\Phi_s=\frac{p^{(\phi_s)}}{p^{(t_s)}},\qquad
\cos\Theta_s=-\frac{p^{(\theta_s)}}{p^{(t_s)}},
\eea
where the frame components are defined in the way of \eqref{eq:FrameComp}.
From these relations we can get
\begin{subequations}
\label{eq:ConservedQuantitiesEmitter}
\bea
\label{eq:HatLambda}
\hat{\lambda}&=&\frac{\cos\Phi_s+v_s}{(\Sigma_s/\Xi_s)\sqrt{\Delta_s}+
\Omega_s\cos\Phi_s+\omega_s v_s},\\
\hat{q}&=&\sqrt{\hat{f}_{\theta}(\frac{\pi}{2})\mp\frac{r_s\cos\Theta}{g}}.
\eea
\end{subequations}

\subsubsection{Near-extremal limit}
In the near-extremal limit, $a=M\sqrt{1-\epsilon^3}$, and for an emitter orbiting on (or near) the ISCO [Eq.~\eqref{eq:SourcePosition}], we can introduce a new quantity $\lambda$ to keep track of the small corrections by expanding Eq.~\eqref{eq:HatLambda}, as
\be
\hat{\lambda}=2M(1-\epsilon\lambda).
\ee

Using the positivity of the kinetic energy in a local frame, we have 
\be
3-\frac{\hat{q}^2-\hat{f}_{\theta}(\theta)}{M^2}<4(1-\lambda\epsilon
+\lambda^2\epsilon^2).
\ee
For convenience, we may introduce a dimensionless shifted Carter constant \cite{porfyriadis2017photon}
\be
q^2=3-\frac{\hat{q}^2}{M^2},
\ee
As mentioned below \eqref{eq:CarterConstant}, we have $Q-f_{\theta}(\theta)\geq0$ for those photons emitted from equatorial plane.
Thus, we have
\be
\label{eq:UpperBoundofQ}
q^2\leq3-\frac{\hat{f}_{\theta}(\theta)}{M^2}.
\ee
Motion in the equatorial plane has $q=\sqrt{3-\hat{f}_{\theta}(\theta)/M^2}$.
We will show in App.~\ref{App:RadialIntegrals} that a light ray originating from this emitter to an observer at far region must also have lower bounds on $q^2$ and we can always have a positive $q$. Together with the upper bound \eqref{eq:UpperBoundofQ}, we can get the range for a specific plasma model.
For the model 1 [Eq.~\eqref{eq:PlasmaCase1}], we may either choose $f_r(r)=0$ and $f_{\theta}(\theta)=\omega_c^2 M^2$, then we have
\be
0<q^2\leq3-\hat{\omega}_c^2;
\ee
or choose $f_r(r)=\omega_c^2M^2$ and $f_{\theta}(\theta)=0$, then we have
\be
\hat{\omega}_c^2<q^2\leq3.
\ee
Note that the range of $q$ may depend on the choice of $f_r$ and $f_{\theta}$, however, the quantity $\hat{q}^2-\hat{f}_{\theta}$ remains unchanged in any case.
For the model 2 [Eq.~\eqref{eq:PlasmaCase2}], we choose $f_r(r)=\omega_c^2 M r$ and $f_{\theta}(\theta)=0$, then we have
\be
(5\hat{\omega}_c^2-\hat{\omega}_c^4)/4<q^2\leq3,\qquad
q^2>\hat{\omega}_c^2.
\ee

\subsection{Observational quantities}
Following Refs.~\cite{cunningham1972optical,cunningham1973optical,Gralla:2017ufe}, we now consider the observational quantities of the orbiting emitter in a plasma: the image position, redshift factor and flux.
These can be expressed in terms of the photon conserved quantities \eqref{eq:ConservedQuantitiesEmitter}.

The apparent position of the image is obtained by plugging Eqs.~\eqref{eq:ConservedQuantitiesEmitter} into Eqs.~\eqref{eq:ScreenCoordinates} with the sign of $\beta$ equal to $s$ (the final vertical orientation of $p_\theta$).

The frequency of photons observed by an observer is shifted from the original frequency when they were emitted from the source. The ``redshift factor" is the ratio between the frequency measured at infinity and measured at local rest frame of the source,
\be
\label{eq:RedshiftGeneral}
g=\frac{\omega_0}{\omega_s}=\frac{\omega_0}{p^{(t_s)}}
=\frac{1}{\gamma}\sqrt{\frac{\Delta_s\Sigma_s}{\Xi_s}}
\frac{1}{1-\Omega_s\hat{\lambda}}.
\ee

The normalized total flux of each image on the observer's screen relative to the comparable `Newtonian flux' in a vacuum is obtained as
\be
\label{eq:FluxGeneral}
\frac{F_o}{F_N}=g^3\frac{\hat{q}M}{\gamma\sin\theta_o}
\sqrt{\frac{\Sigma_s\Delta_s}
{\Xi_s\Theta(\theta_o)\Theta(\theta_s)\mathcal{R}(r_s)}}\left|
\det\frac{\partial(B,A)}
{\partial(\hat{\lambda},\hat{q})}\right|^{-1},
\ee
where we have defined
\subbe
\label{eq:DefAB}
\bea
\label{eq:DefA}
A&\equiv& I_r+b\tilde{I}_r-G^{m,s}_{\theta}\pm M\int^{\theta_s}_{\pi/2}\frac{d\theta}{\sqrt{\Theta(\theta)}},\\
B&\equiv& J_r+b\tilde{J}_r+\frac{\hat{\lambda}G^{m,s}_{\phi}-\Omega_sa^2G^{m,s}_t}
{M}.
\eea
\subee

\section{Integrals and lens equations}
\label{App:Integrals}
In this appendix, we write down the integrals appearing in the ``lens equations" \eqref{eq:LensEqns} for plasma models that satisfy the separation condition \eqref{eq:SeparationCondition} with some undetermined functions $f_r$ and $f_{\theta}$. Note that the functions $f_r$ and $f_{\theta}$ are appearing in the radial integral and angular integral, respectively. We compute those integrals for some specific choices of the functions $f_r$ and $f_{\theta}$. The results of the lens equations for the power-law-like models \eqref{eq:PlasmaCase1} and \eqref{eq:PlasmaCase2} are obtained by plugging in the corresponding integrals at the same time.
\subsection{Radial integrals}
\label{App:RadialIntegrals}
\label{app:MAEandRadialIntegral}
The radial integrals appearing in the ``lens equations" \eqref{eq:LensEqns} are defined as \cite{Gralla:2017ufe}
\begin{subequations}
\bea
\label{eq:IntergralIr}
I_r&=&M \int^{r_o}_{r_s}\frac{dr}{\sqrt{\mathcal{R}(r)}},\quad
\tilde{I}_r=2M\int^{r_s}_{r_{\text{min}}}\frac{dr}{\sqrt{\mathcal{R}(r)}},\\
\label{eq:IntergralJr}
J_r&=&\int^{r_o}_{r_s}\frac{\mathcal{J}_r}{\sqrt{\mathcal{R}(r)}}dr,\quad
\tilde{J}_r=2\int^{r_s}_{r_{\text{min}}}\frac{\mathcal{J}_r}{\sqrt{\mathcal{R}(r)}}dr,
\eea
\bea
\label{eq:MathcalJr}
\mathcal{J}_r&=&\frac{1}{\Delta}\Big[a(2Mr-a\hat{\lambda})-\Omega_sr
\big(r^3+a^2(r+2M)\nn\\
&&-2aM\hat{\lambda}\big) \Big],
\eea
\end{subequations}
where the radial potential $\mathcal{R}(r)$ is defined in \eqref{eq:RadialPotential} and $r_{\text{min}}$ is the largest (real) root of it. These equations are valid when $r_{\text{min}}<r_s$, which is always true for light that can reach infinity.

In the near-extremal limit, we have expansions \eqref{eq:NearExtremalExpansion}.
We will perform these integrals analytically to the leading order in $\epsilon$ by using matched asymptotic expansions (MAE) for light rays connecting a source to an observer. We work in the dimensionless radial coordinates $R$ [defined in Eq.~\eqref{eq:DimensionlessR}] and split each of the integrals into two pieces by introducing a scaling of $\epsilon^p$ ($\epsilon\ll\epsilon^p\ll1$) with a constant $p\in(0,1)$.

Take $I_r$ for example, under this regime we have
\be
\label{eq:SplitIntegral}
I_r\approx M^2\int^{\epsilon^p C}_{\epsilon\bar{R}}\frac{dR}{\sqrt{\mathcal{R}_n}}+M^2\int^{R_o}_{\epsilon^p C}\frac{dR}{\sqrt{\mathcal{R}_f}},
\ee
where $\mathcal{R}_n$ and $\mathcal{R}_f$ are the leading terms of the expansions in the near horizon region $R\sim\epsilon$ and in the far region $R\sim1$, respectively, and $C$ is a positive constant. Note that scaling of $R\sim\epsilon^pC$ is in the overlap region.

For $f_r(r)=M^{2-k}\omega_c^2r^k$ ($0\leq k\leq2$), we have
\subbe
\label{eq:RadialPotentialExpansions}
\bea
\mathcal{R}_n(R\sim\epsilon)&=&M^4\epsilon^2\big[(q^2
-\hat{\omega}_c^2)x^2+4\lambda(2x+\lambda)\big],\\
\mathcal{R}_f(R\sim 1)&=&M^4\big[q^2+4R+R^2-(1-R)^k\hat{\omega}_c^2\big],
\eea
\subee
where we have introduced $x=R/\epsilon$. At every point of a photon trajectory that originates in the NHEK region and reaches to the far region, one must have non-negative potential $\mathcal{R}(r)\geq0$. To guarantee that this condition is hold in the near region, we should take $q^2>\hat{\omega}_c^2$; to guarantee that this condition is hold in the far region, we should take $q^2>\hat{\omega}_c^2$ for $k=0$ and $q^2>(5\hat{\omega}_c^2-\hat{\omega}_c^4)/4$ for $k=1$.

With the expansions \eqref{eq:RadialPotentialExpansions}, we can analytically compute the two pieces of integrals in Eq.~\eqref{eq:SplitIntegral}, respectively, and the complete integral is obtained by adding up them. Moreover, the integrals $\tilde{I}_r$, $J_r$ and $\tilde{J}_r$ can be computed in a similar way.

Next, we list the results for $\hat{f}_r(r)=\hat{\omega}_c^2Mr$, as follows,
\subbe
\bea
I_r&=&\frac{1}{\tilde{q}}
\log\Big[\frac{4\tilde{q}^4R_o}{\big(\tilde{q}^2
+(2-\frac{\hat{\omega}_c^2}{2})R_o+\tilde{q}
D_o\big)\big(\tilde{q}D_s+\tilde{q}^2\bar{R}
+4\lambda\big)}\Big]\nn\\
&&-\frac{\log\epsilon}{\tilde{q}}+\mathcal{O}(\epsilon),\\
\tilde{I}_r&=&\frac{1}{\tilde{q}}\log\Big[\frac{
\big(\tilde{q}D_s+\tilde{q}^2\bar{R}
+4\lambda\big)^2}{4(4-\tilde{q}^2
)\lambda^2}\Big]+\mathcal{O}(\epsilon),\\
J_r&=&\log\Bigg[\frac{\bar{R}
(2+\tilde{q})(2-\frac{\hat{\omega}_c^2}{2}
+\tilde{q})^{1+\frac{1}{4}\hat{\omega}_c^2}}{(D_s+2\bar{R}+2\lambda)
(2-\frac{\hat{\omega}_c^2}{2}+D_o+R_o)^{1+\frac{1}{4}\hat{\omega}_c^2}}
\Bigg]\nn\\
&&-\frac{7}{2}I_r+\frac{3}{8\lambda}(D_s-\tilde{q}\bar{R})+
\frac{1}{2}(\tilde{q}-D_o)
 +\mathcal{O}(\epsilon),\\
\tilde{J}_r&=&-\frac{7}{2}\tilde{I}_r-\frac{3}{4}\frac{D_s}{\lambda}+
\log\Big[\frac{(D_s+2\bar{R}+2\lambda)^2}{(4-\tilde{q}^2)\bar{R}^2}\Big]
+\mathcal{O}(\epsilon),
\eea
\subee
where $\tilde{q}=\sqrt{q^2-\hat{\omega}_c^2}$, and
\subbe
\label{eq:DoDsCase2}
\bea
D_s&=&\sqrt{\tilde{q}^2\bar{R}^2+8\lambda\bar{R}+4\lambda^2},\\
D_o&=&\sqrt{\tilde{q}^2+(4-\hat{\omega}_c^2)R_o+R_o^2}.
\eea
\subee

Note that for $\hat{\omega}_c=0$, these give the results for $\hat{f}_r(r)=0$.

For $\hat{f}_r(r)=M^2\hat{\omega}_c^2$, the expansions \eqref{eq:RadialPotentialExpansions} are similar as those for $\hat{f}_r(r)=0$ up to a replacement of $q\rightarrow\sqrt{q^2-\hat{\omega}_c^2}$. Thus, the final results of the integrals are obtained by including this replacement in those for $\hat{f}_r(r)=0$.

\subsection{Angular integrals}
\label{App:AngularIntegrals}
The angular integrals appearing in the ``lens equations" \eqref{eq:LensEqns} are defined as \cite{Gralla:2017ufe}
\begin{align}
\label{eq:AngleIntegralMS}
	G^{m,s}_i=
	\begin{cases}
		\hat{G}_i\qquad\qquad&m=0,\\
		mG_i-s\hat{G}_i\qquad&m\ge1,
	\end{cases}
	\qquad
	i\in\cu{t,\theta,\phi},
\end{align}
with
\be
G_i=M\int^{\theta_+}_{\theta_-}g_i(\theta)d\theta,\quad
\hat{G}_i=M\int^{\pi/2}_{\theta_o}g_i(\theta)d\theta,
\ee
and
\be
g_{\theta}=\frac{1}{\sqrt{\Theta(\theta)}},\quad
g_{\phi}=\frac{\csc^2\theta}{\sqrt{\Theta(\theta)}},\quad
g_{t}=\frac{\cos^2\theta}{\sqrt{\Theta(\theta)}},
\ee
where $\Theta(\theta)$ is the angular potential defined in \eqref{eq:AngularPotential} with an arbitrary function $f_{\theta}(\theta)$ in it
and $\theta_{\pm}$ are roots of it.

We will perform the integrals for $f_{\theta}(\theta)=\omega_c^2 M^2$. In this case, the angular potential can be written as
\be
\Theta(u)=\hat{q}^2-\hat{\omega}_c^2M^2+u\big[a^2-\hat{\lambda}^2(1-u)^{-1}\big],
\ee
which are similar as those in the vacuum Kerr case up to the replacement:
$
\hat{q}^2
\rightarrow
\hat{q}^2-\hat{\omega}_c^2M^2.
$
Thus, following Ref.~\cite{Gralla:2017ufe} we obtain the results of the angular integrals in the near-extremal regime, as follows
\begin{subequations}
\bea
G_{\theta}&=&\frac{2}{\sqrt{-\mathcal{I}_-}}K\Bigg(
\frac{\mathcal{I}_+}{\mathcal{I}_-}\Bigg)+\mathcal{O}(\epsilon),\\
\hat{G}_{\theta}&=&\frac{1}{\sqrt{-\mathcal{I}_-}}F\Bigg(\Psi_o\Big|
\frac{\mathcal{I}_+}{\mathcal{I}_-}\Bigg)+\mathcal{O}(\epsilon),\\
G_{\phi}&=&\frac{2}{\sqrt{-\mathcal{I}_-}}\Pi\Bigg(\mathcal{I}_+\Big|
\frac{\mathcal{I}_+}{\mathcal{I}_-}\Bigg)+\mathcal{O}(\epsilon),\\
\hat{G}_{\phi}&=&\frac{1}{\sqrt{-\mathcal{I}_-}}\Pi
\Bigg(\mathcal{I}_+;\Psi_o\Big|
\frac{\mathcal{I}_+}{\mathcal{I}_-}\Bigg)+\mathcal{O}(\epsilon),\\
G_{t}&=&-\frac{4\mathcal{I}_+}{\sqrt{-\mathcal{I}_-}}E^{\prime}\Bigg(
\frac{\mathcal{I}_+}{\mathcal{I}_-}\Bigg)+\mathcal{O}(\epsilon),\\
\hat{G}_{t}&=&-\frac{2\mathcal{I}_+}{\sqrt{-\mathcal{I}_-}}
E^{\prime}\Bigg(\Psi_o\Big|
\frac{\mathcal{I}_+}{\mathcal{I}_-}\Bigg)+\mathcal{O}(\epsilon),
\eea
\end{subequations}
where $E^{\prime}(\phi|m)=\partial_m E(\phi|m)$ and
\be
\Psi_o=\arcsin\sqrt{\frac{\cos^2\theta_o}{\mathcal{I}_+}}, 
\ee
and
\be
\mathcal{I}_{\pm}=\frac{\bar{q}^2}{2}-3\pm\sqrt{12-(2\bar{q})^2
+\Big(\frac{\bar{q}^2}{2}\Big)^2},
\ee
with
\be
\bar{q}=\sqrt{q^2+\hat{\omega}_c^2}.
\ee
Furthermore, $F(\phi|m)$, $E(\phi|m)$, $\Pi(n;\phi|m)$ are the incomplete elliptic integrals of the first, second and third kind, respectively, and $K(m)=F(\frac{\pi}{2} |m)$, $E(m)=E(\frac{\pi}{2}|m)$, $\Pi(n|m)=\Pi(n;\frac{\pi}{2}|m)$ are the corresponding complete elliptic integrals.

Note that for $\hat{\omega}_c=0$, these give the results for $\hat{f}_{\theta}(\theta)=0$.
\subsection{Lens equations for the two plasma models}
In Sec.~\ref{Subsec:NearExtremalSolution}, we have solved the lens equations for the power-law-like plasma models 1 and 2 [Eqs.~\eqref{eq:PlasmaCase1} and \eqref{eq:PlasmaCase2}] by choosing $f_r(r)=0$, $f_{\theta}(\theta)=\omega_c^2M^2$ for model 1 and $f_r(r)=\omega_c^2Mr$, $f_{\theta}(\theta)=0$ for model 2. In order to compare the len equations for these two models, we may equivalently choose $f_r(r)=\omega_c^2M^2$, $f_{\theta}(\theta)=0$ for model 1 instead. Thus, the difference of the lens equations between these two models are only imposed in the radial integrals with the function $f_r(r)$ taken the form of $f_r(r)=M^{2-k}\omega_c^2r^k$. Moreover, from the expansions \eqref{eq:RadialPotentialExpansions} we see that the near-horizon piece of the radial integrals\linebreak for the two models are exactly the same, while even though the far region piece contains differences among these two models, the plasma densities are small in that region since they scale like $1/r^h$ ($h$=2 and 1, respectively).

\end{appendix}


\providecommand{\href}[2]{#2}\begingroup\raggedright\endgroup

\end{document}